\documentstyle[amssymb,pre,twocolumn,aps]{revtex}
\begin{document}
\title{Transition Between Ground State and Metastable States in Classical 2D Atoms}
\author{Minghui Kong, B. Partoens, and F. M. Peeters\cite{pet}}
\address{{\it Departement Natuurkunde, Universiteit Antwerpen (UIA)}\\
{\it Universiteitsplein 1, B-2610 Antwerpen, Belgium}}
\maketitle

\begin{abstract}
Structural and static properties of a classical two-dimensional (2D) system
consisting of a finite number of charged particles which are laterally
confined by a parabolic potential are investigated by Monte Carlo (MC)
simulations and the Newton optimization technique. This system is the
classical analog of the well-known quantum dot problem. The energies and
configurations of the ground and all metastable states are obtained. In
order to investigate the barriers and the transitions between the ground and
all metastable states we first locate the saddle points between them, then
by walking downhill from the saddle point to the different minima, we find
the path in configurational space from the ground state to the metastable
states, from which the geometric properties of the energy landscape are
obtained. The sensitivity of the ground-state configuration on the
functional form of the inter-particle interaction and on the confinement
potential is also investigated.

\end{abstract}

\section{INTRODUCTION}

Wigner suggested in 1934 that a liquid to solid phase transition should
occur in a three-dimensional (3D) Fermi system at low densities \cite{1}.
The quest for the observation of such a Wigner crystal has been the object
of very intense and not stagnant work. After the first discovery of Wigner
crystallization of electrons on the surface of liquid helium \cite{2}, there
has been considerable theoretical and experimental progress in the study of
the mesoscopic system consisting of a finite number of charged particles,
which are laterally confined by a parabolic potential and repel each other
through a Coulomb potential. This system is the classical analog of the
well-known quantum dot problem. These quantum dots are atomic-like
structures which have interesting optical properties and may be of interest
for single electron devices. These systems and their configurations have
been observed experimentally, and are important in solid-state physics,
plasma physics as well as in atomic physics. The classical approach is valid
for quantum dots in high magnetic fields where the kinetic energy of the
electrons is quenched, or for other classical systems, such as laser cooled
ions in a trap \cite{3} which are realized by electric and magnetic fields,
trapped ions cooled by laser techniques \cite{4}, ions in a radio-frequency
(RF) trap (Paul trap) \cite{5,6} or a Penning trap \cite{7,8,9} which can
also serve as an illustration of 3D Coulomb clusters \cite{9.1,9.2}. Very
large Coulomb clusters have been created recently in strongly coupled RF
dusty plasmas \cite{10,11,12} which are like a two-dimensional (2D) layered
system. Examples of 2D Coulomb clusters are electrons on the surface of
liquid helium \cite{13} and electrons in quantum dots \cite{14}. The vortex
clusters in an isotropic superfluid \cite{15}, vortices in superfluid He$%
^{4} $\cite{16,17}, vortices in a Bose-Einstein condensate stirred with a
laser beam \cite{18} and in superconducting grains \cite{19} have many
common features with those of 2D charged particles \cite{20}. Colloidal
particles dissolved in water \cite{20.1,20.2} and placed between two glass
plates are another example of an experimental system where classical
particles exhibit Wigner crystallization \cite{23}. Very recently,
macroscopic 2D Wigner islands, consisting of charged metallic balls above a
plane conductor were studied and ground state, metastable states and saddle
point configurations were found experimentally \cite{24}.

In a finite system there is a competition between the bulk triangular
lattice and the circular confinement potential which tries to force the
particles into a ring like configuration. Those configurations were
systematically investigated in Ref. \cite{25} and a Mendeleev-type of table
for these classical atomic-like structures was constructed. The spectral
properties of the ground state configurations were presented in Refs. \cite
{26,27} and generalized\ to screened Coulomb \cite{30,31} and logarithmic 
\cite{30,32,33,35} interparticle interactions.

In the present paper we want to go one major step further and calculate not
only all the different metastable states but also the saddle points between
those local energy minima and the path followed by the particles to transit
between those energy minima. The present work is motivated by recent
experimental work \cite{24} where it was found that: i) some of the
configurations did not agree with the previous theoretical published one,
and ii) they were able to observe some of the saddle points which are the
key configurations for transition between different stable (ground or
metastable) states. Therefore, we also investigated the stability of the
ground state configurations against the functional form of the confinement
potential and the exact form of the inter-particle interaction potential.

The present paper is organized as follows. In Sec. 2, we describe the model
system. In Sec. 3, our numerical technique, used to obtain the ground and
metastable states, is outlined. The technique we used to find the saddle
point is similar to the Cerjan-Miller algorithm \cite{35.5}. After the
saddle points are found, we connect the saddle point to the global minimum
or a local minimum by the `walking downhill' method. Sec. 4 is devoted to
the structural and static properties of the ground and metastable states for 
$N=1\thicksim 40$. The configurations are analyzed and compared with
available experimental data and the results of previous theoretical
approaches. The phase diagram for 9 and 16 particles in the ground state
with different functional forms of confinement potential and interparticle
interaction is also calculated. The discussion on the saddle point is
presented in Sec. 5, and the connecting path from the ground state to the
metastable states is found, and we investigate the completely geometric
properties of the energy landscape. Our conclusions are presented in Sec. 6.

\section{Model system}

The model system consists of identical charged particles interacting through
a Coulomb repulsive interacting and moving in a 2D plane where they are
confined by a parabolic potential 
\begin{equation}
H=\frac{q^{2}}{\varepsilon }\sum_{i>j}\frac{1}{\left| \overrightarrow{r_{i}}-%
\overrightarrow{r}_{j}\right| }+\sum_{i}V(\overrightarrow{r}_{i}).
\end{equation}
The confinement potential $V(\overrightarrow{r})=\frac{1}{2}m^{\ast }\omega
_{0}^{2}r^{2}$ is taken circular symmetric and parabolic, where $m^{\ast }$
is the effective mass of the particles, $q$ is the particle charge, $\omega
_{0}$ is the radial confinement frequency and $\varepsilon $ is the
dielectric constant of the medium the particles are moving in. Note that for
the quantum dot problem an additional term appears in Eq. (1) which is the
kinetic energy of the particles which is absent in our statical classical
problem. Here the motion of the particles is restricted to the $(x,y)$
plane. To exhibit the scaling of the system, we introduce the characteristic
scales in the problem: $r_{0}=(2q^{2}/m\epsilon \omega _{0}^{2})^{1/3}$ for
the length and $E_{0}=(m\omega _{0}^{2}q^{4}/2\epsilon ^{2})^{1/3}$ for the
energy. After the scaling transformations ($r\rightarrow
r/r_{0},E\rightarrow E/E_{0}$), the Hamiltonian can be rewritten in a simple
dimensionless form as

\begin{equation}
H=\sum_{i>j}\frac{1}{\left| \overrightarrow{r_{i}}-\overrightarrow{r}%
_{j}\right| }+\sum_{i}V(\overrightarrow{r_{i}}),
\end{equation}
with $V(\overrightarrow{r})=x^{2}+y^{2}$ and which only depends on the
number of particles $N$. The numerical values for the parameters $\omega
_{0},r_{0},E_{0}$ for some typical experimental systems were given in Ref. 
\cite{25}.

\section{Numerical approach}

Most of the previous works have treated the quantum mechanical problem of a
small number of electrons. In the present paper, we consider only the
classical system. Although a classical approach for the description of the
behavior of electrons in quantum dots, in principle, is not applicable, it
is possible that certain features of the classical system may survive in a
quantum system.

Due to the presence of confinement energy and electron-electron Coulombic
interaction, a complete description of the cluster system is complicated and
can't be obtained analytically. The Monte Carlo simulation technique \cite
{36} is relatively simple and rapidly convergent and it provides a reliable
estimation of the total energy of the system in cases when relatively small
number of Metropolis steps is sufficient. However, the accuracy of this
method in calculating the explicit states is poor in certain cases. It
becomes more difficult for clusters with a large number of particles, which
have significantly more metastable states. To circumvent this problem we
employ the numerical technique of Newton optimization which was outlined and
compared with the standard Monte Carlo technique in Ref. \cite{26}. In this
way, we are able to obtain not only the ground state but also metastable
states. It also yields the eigenfrequencies and the eigenmodes of the ground
state configuration. Now only a small number of calculation steps is needed
to obtain the same accuracy. Moreover, using the modified Newton approach,
we can explore the stability of the system in its ground-state configuration
through its spectrum.

By studying the characteristics of the energy landscape and the energy
barrier between the different local minima, we are able to find the saddle
point configurations which are very important and are the key configurations
for transition between different stable states. The technique we used to
find the saddle point is explained in more detail in Ref. \cite{37}, and is
similar to the Cerjan-Miller algorithm \cite{35.5}. After the saddle points
are found, we connect the saddle point to the global minimum or a local
minimum by the `walking downhill' method. In this algorithm the direction of
the steepest gradient is followed to force the system to transit from the
saddle point state to the local minimum state. Which minimum is finally
reached depends on the initial step, therefore we repeat this procedure
several times to determine both minima which the saddle point state
connects. Thus the connecting path followed by the particles to transit
between those energy minima is found, from which the geometric properties of
the energy landscape are obtained.

\section{Ground state and metastable state}

In Table \ref{Tab1} we list for $N$=1, 2,...,40 the energy per particle $E/N$
in the ground state and in the metastable states, where we also list the
energy difference with the ground state $\Delta E/N.$ The configuration is
indicated by the number of particles in the different rings, the position of
the center of the ring and the radius of the different rings, the width of
the ring which is defined as the difference of the maximum radius and
minimum radius in the same ring, and the energy of the lowest three normal
mode frequencies of the ground state are also given in Table I. This table
is rather exhaustive and should be compared with a similar one published in
Ref. \cite{33} for a logarithmic interacting system.

For different values of $N$ there exist different possible values for $E/N$
which are nothing else than the metastable states. The difference in energy
between the metastable and the ground state is given in the third column and
the corresponding configuration in the fourth column. Note that with
increasing $N$ the number of metastable configurations increases and in
general (but not always) the widths of the rings for metastable
configurations are larger and the central ring/particle is not exactly
located in the center of the parabolic potential well. For sufficiently
large $N$, the simple ring structure gradually disappears in the center and
the triangular Wigner lattice appears. There is a competition between two
types of ordering: ordering into a triangular-lattice structure (Wigner
lattice) and ordering into a shell structure, which leads to clusters with
interesting self-organized patterns which show concentric shells at small $N$
and hexagonal cores surrounded by circular outer shells at large $N$.

The lowest non-zero normal mode frequency is a measure for the stability of
the ground state, it tells us how easy or difficult it is to deform this
state. Therefore, intuitively we would expect that the value of this
frequency would be correlated to $\Delta E/N$, the energy difference between
the first metastable and the ground state. Those values are plotted in Fig. 
\ref{Fig1} as function of $N$. Notice that there exist such a correlation in
general, but that this is not true for all $N$-values, e.g. for $%
N=12,18,19,20,21,30$ there is no correlation.

The rings have sometimes a finite width which are shown in Fig. \ref{Fig2}
as a function of $N$. Notice that the widths fall into three bands; i) width 
$\leqslant $ 0.003 which is practically a perfect ring, ii) width $\thicksim 
$ 0.02, and iii) width $\thicksim $ 0.5. Usually, but not always, the outer
ring has the largest width. The width of the rings increases with increasing 
$N$ and at the same time the widest ring becomes often the next to outer
ring.

We compare our ground state configuration with available experimental data 
\cite{24} and the results of previous theoretical approaches \cite
{25,30,32,33}. For very small number of particles ($N<16$), all theoretical
and experimental results for the ground state configurations are the same
expect for $N$=9 and 15 whatever kind of interparticle interaction. The
experimental observation \cite{24} for the ground state of 9 particles is
(1,8) and for 15 particles it is (4,11), which compares to our result (2,7)
and (5,10), respectively. For $17<N<30,$ the experimental result and all the
calculated patterns present three shells. Our result differs with the
experimental data of Ref. \cite{24} for $N$= 17, 20, 22, 24, 25, 27-30.
Because of the discrepancy between some of the experimental configurations
and the `numerical exact' theoretical ground state configurations it is
possible that experimentally the inter-particle interaction is not exactly a
Coulombic potential and the confinement potential is not purely quadratic.
Therefore, we investigated the effect of such deviations of these potentials
on the ground state configuration. As an example we took $N$=9 and use
confinement potentials $V\thicksim r^{n}$ and for the inter-particle
interaction $V\thicksim r^{-m}$. The resulting phase diagram is shown in
Fig. \ref{Fig3}. Notice that, depending on the values of $n$ and $m,$ the
system can be either in the (1,8) or the (2,7) configuration. For the
harmonic confined Coulomb interacting system i.e. ($n,m$) = (2,1) the system
is in the (2,7) configuration but from the phase diagram it is clear that if
we change the confinement potential slightly and make it more steep up to $%
n\geqslant $2.2 the configuration (1,8) becomes the ground state. The
experimentally determined ground state configuration for 9 particles was
(1,8) [24].

There is also a difference with the experimental data and our results for $N$%
=16 particles. Therefore, we did the same investigation and present the
phase diagram in Fig. \ref{Fig4}. Notice that the harmonic confined Coulomb
interacting system, i.e. $(n,m)$ = (2,1) is again close to the phase
boundary between the configuration (1,5,10) and (5,11). This is probably the
explanation why the experimental configuration\cite{24} differs from our
simulation results, since it is hard to guarantee that ($n,m$) is exactly
(2,1) during the experiment.

Notice that for both $N$=9 and $N$=16 the metastable configuration has an
energy very close to the one of the ground state, the difference is less
than 0.2$\%$. These metastable configurations correspond indeed with the
experimentally observed ones. Consequently, an alternative explanation for
the difference with the experiment is that the experimental configuration
got stuck in the metastable configuration.

\section{Saddle points}

Between metastable states and the ground state there are potential barriers.
The system will prefer to transfer over the lowest potential barrier, which
is the saddle point configuration between these energy minima, in order to
transit from one stable configuration to the other. We plot in Fig. \ref
{Fig5} the trajectories of the particles for the $N$=5 system making a
transition from the ground state (5) to the metastable state (1,4) and the
saddle point connecting them. The trajectories of the particles can also be
obtained by moving one of the particles to the center of the system.

For 6 particles, the ground state (1,5) and the metastable state (6),
corresponding to the hexagonal configuration, are obtained. Moreover, the
unstable equilibria associated to saddle point configurations are also
obtained, and the energy landscape is shown schematically\ in Fig. \ref{Fig6}%
. There are two saddle points for this case, one of them is very close to
the metastable state in both energy and configuration, and will therefore be
hard to see experimentally \cite{24}. In Fig. \ref{Fig6}, the insets show
the arrangement of the particles for the different states. Using the
`walking downhill method', we found the central particle slowly moving to
the periphery of the cluster. We would like to stress that the configuration
with 6 particles on a perfect ring is a saddle point state in contrast to
the claim\ made in Ref. \cite{38}. This can be understood from the following
simple model calculation: if 3 particles are placed on a circle with radius $%
A$, on the corners of an equilateral triangle, and the other 3 particles on
another equilateral triangle's corners with radius $B$ rotated over $60^{0}$%
, the energy is 
\begin{equation}
E(c)=\frac{9}{2}\left( \frac{1+c^{2}}{36}\right) ^{1/3}(\frac{1}{1+c}+\frac{%
1+c}{\sqrt{3}c}+\frac{2}{\sqrt{1-c+c^{2}}})^{2/3}
\end{equation}
where $c=\frac{A}{B}$. This function is shown in Fig. \ref{Fig7}. It is
clear that the perfect circle configuration i.e. $c=\frac{A}{B}=1$ is a
saddle point, and that the minimum is obtained if 3 particles move a bit to
the center, and the other 3 particles move away from the center (see the
insets in Fig. \ref{Fig7}). Both shown metastable states are just connected
by a rotation over $120^{0}$. The two minima in Fig. \ref{Fig7} correspond
to the same configuration in which inner and outer ring are interchanged.
Comparing our results with the Fig. 2 (`$N=6$: ground state, saddle point
configuration and the hexagonal metastable state') of Ref. \cite{24}, we see
that the other saddle point is observed experimentally.

A list of the saddle point energies up to 20 particles is given in Table \ref
{Tab2}. From this table, we notice that there is only one saddle point state
for $N=3,4,5$ particles. But, on the other hand it is well-known that there
are $(k-1)$ saddle points when there are $k$ minima. For $N$=3 and 4 one
saddle point is found, although there is no metastable configuration. The
reason is that the saddle point state connects two equilateral ground state
configurations which can be obtained from each other by a simple rotation.
For the simple case of 3 particles, we show the energy surface and the
corresponding configurations schematically in Fig. \ref{Fig8}. Notice that
there are always more saddle points than minima for $N$%
%TCIMACRO{\TEXTsymbol{>}}%
%BeginExpansion
\mbox{$>$}%
%EndExpansion
6. With increasing the number of particles, more saddle point states are
obtained and the energy landscape gets more complicated. For example for 9
particles, we obtain three saddle points and one metastable state. The
results for the trajectories and energy landscape are shown in Fig. \ref
{Fig9}. Again, the ground state configurations corresponding with the black
and the white dot are connected by a simple rotation, i. e. a symmetry
operation.

\section{Conclusion}

We presented the results of a numerical calculation of the configurations of
the ground and all metastable states and their energies, the system
consisting of classical 2D charged particles that are confined in a
parabolic confinement potential for $N$=1,...,40. These artificial atoms
undergo configurational changes when the system transits from the ground
state to the different metastable states, or between the different
metastable states. Such transitions move through the lowest energy barrier
connecting those states, i. e. through a saddle point. The connecting path
from the ground state to all metastable states is found and the geometric
properties of the energy landscape were discussed.

Sensitivity of the configuration on the form of the confinement potential
and the interparticle interaction is investigated and a phase diagram was
obtained. This sensitivity on e.g. the form of the confinement potential is
probably the explanation why the experimental configuration\cite{24} differs
from our simulation results.

\section{Acknowledgments}

B. Partoens is a post-doctoral researcher of the Flemish Science Foundation
(FWO-Vlaanderen). Stimulating discussions with Dr. J. Shi and M. Milo\v{s}%
evi\'{c} are gratefully acknowledged. This work is supported by the Flemish
Science Foundation (FWO-VI), the Belgian Inter-University Attraction Poles
(IUAP-VI), the ``Onderzoeksraad van de Universiteit Antwerpen''(GOA), the EU
Research Training Network on ``Surface Electrons on Mesoscopic Structures'',
and INTAS.

\bigskip

Tables

\begin{table}[tbp]
\caption{The ground state and the metastable states for $N$=1,..., 40
Coulombic particles confined in a 2D parabolic well. We give the energies ($%
E/N$), $\Delta E/N,$ the shell structure ($N_{1},N_{2},...$), the radius and
width of the shell, and the lowest three normal mode frequencies of the
ground state configuration. }
\label{Tab1}
\end{table}

\begin{table}[tbp]
\caption{The energies of the ground state, the metastable states and the
saddle point states for different number of particles ($N$).}
\label{Tab2}
\end{table}

Figure captions

\begin{figure}[tbp]
\caption{The lowest eigenfrequency and $\Delta E/N$ as function of the
number of particles.}
\label{Fig1}
\end{figure}

\begin{figure}[tbp]
\caption{The width of the different shells (logarithmic scale) as function
of the number of particles.}
\label{Fig2}
\end{figure}

\begin{figure}[tbp]
\caption{The phase diagram for the ground state of 9 particles. The
dependence on the form of the confinement potential and the interparticle
interaction is shown.}
\label{Fig3}
\end{figure}

\begin{figure}[tbp]
\caption{The phase diagram for the ground state of 16 particles. The
dependence on the form of the confinement potential and the interparticle
interaction is shown.}
\label{Fig4}
\end{figure}

\begin{figure}[tbp]
\caption{The trajectories of the particles making a transition from the
ground state to the metastable state and the saddle point connecting them
for 5 particles.}
\label{Fig5}
\end{figure}
\begin{figure}[tbp]
\caption{The energy landscape and transition between the ground state to the
metastable states for 6 particles.}
\label{Fig6}
\end{figure}
\begin{figure}[tbp]
\caption{Part of the energy landscape and corresponding configurations near
the metastable state for 6 particles. }
\label{Fig7}
\end{figure}

\begin{figure}[tbp]
\caption{Schematic view of the energy surface and projection of the energy
and the corresponding configurations for 3 particles.}
\label{Fig8}
\end{figure}
\begin{figure}[tbp]
\caption{The energy landscape and transition from ground state to metastable
states for 9 particles.}
\label{Fig9}
\end{figure}

\end{document}